\documentclass[12pt]{iopart}

\usepackage{graphicx}

\begin{document}

\title[]{Ultrasmall divergence of laser-driven ion beams from nanometer thick foils}

\author{J\ H Bin$^{1,2}$,  W J Ma$^{1,2}$\footnote[1]{Author to whom any correspondence should be addressed}, K Allinger$^{1,2}$, H Y Wang$^{1,2,3}$, D Kiefer$^{1,2}$, S Reinhardt$^{1}$, P Hilz$^{1}$, K Khrennikov$^{1,2}$, S Karsch$^{1,2}$, X Q Yan$^{3}$, F Krausz$^{1,2}$, T Tajima$^{1}$, D Habs$^{1,2}$, J Schreiber$^{1,2}$\footnotemark[1]}

\address{$^1$Fakult$\ddot{a}$t f$\ddot{u}r$ Physik, Ludwig-Maximilians-Universit$\ddot{a}$t M$\ddot{u}$nchen, D-85748 Garching, Germany}
\address{$^2$Max-Planck-Institut f$\ddot{u}$r Quantenoptik, D-85748 Garching, Germany}
\address{$^3$State Key Laboratory of Nuclear Physics and Technology,and Key Lab of High Energy Density Physics Simulation, CAPT, Peking University, Beijing 100871, China}

\ead{\mailto{wenjun.ma@physik.uni-muenchen.de}, \mailto{joerg.schreiber@mpq.mpg.de}}

\begin{abstract}
We report on experimental studies of divergence of proton beams from nanometer thick diamond-like carbon (DLC) foils irradiated by an intense laser with high contrast. Proton beams with extremely small divergence (half angle) of 2$^\circ$ are observed in addition with a remarkably well-collimated feature over the whole energy range, showing one order of magnitude reduction of the divergence angle in comparison to the results from $\mathrm{\mu m}$ thick targets. We demonstrate that this reduction arises from a steep longitudinal electron density gradient and an exponentially decaying transverse profile at the rear side of the ultrathin foils. Agreements are found both in an analytical model and in particle-in-cell simulations. Those novel features make $\mathrm{nm}$ foils an attractive alternative for high flux experiments relevant for fundamental research in nuclear and warm dense matter physics.   

\end{abstract}

\maketitle

\section{Introduction}

The emission of highly energetic ions from solid targets irradiated by intense laser pulses has attracted great attention over the past decades \cite{experiments}. The short time scale on which the acceleration occurs along with the small source size enable extremely high ion densities in the MeV bunches which could be superior for specific applications \cite{application}. However, such high density is only maintained close to the source, it drops quickly due to the angular spread of few tens of degrees \cite{divergence,divcontrol}. Such large angles lead to large losses using magnetic quadrupoles \cite{qp}, complicate the beam transport and therefore trigger investigations on sophisticated transportation schemes such as pulsed solenoid\cite{solenoid} and laser driven micro lenses \cite{secondp}. Meanwhile, shaped lens target \cite{lenstarget}, droplets \cite{droplets} and curved target \cite{curved} have been used to manipulate the ions angular distribution. Those approaches were mostly based on target normal sheath acceleration (TNSA) \cite{tnsa} with $\mathrm{\mu m}$ thick targets. Acceleration fields are built at the target rear by the hot electrons generated at the front side of the targets. The divergence of the ions strongly depends on the electron density and phase space distribution of the electrons behind the target, which is initially related to the laser profile and then disturbed during the transportation through the targets \cite{elespatial}.

Recently, ultrathin foils with thickness down to $\mathrm{nm}$ scale have been investigated experimentally \cite{thinfoil,thinfoil2,thinfoil3}, enabled by the improvement on the laser temporal contrast. A divergent ion beam with opening half angle of roughly 10$^\circ$ from 50 $\mathrm{nm}$ ultrathin foils can be inferred from \cite{thinfoil2}. Proton beams with divergence of 5-6$^\circ$ have been observed from 800 $\mathrm{nm}$ CH targets \cite{thinfoil3}. These results imply that thinner foils can generate much more collimated ion emission as compared to $\mathrm{\mu m}$ thick targets. 

In this letter, we present the first detail study of the divergence of proton beams accelerated from 5-20 $\mathrm{nm}$ thick Diamond-Like-Carbon (DLC) foils \cite{dlc}. Divergences as low as 2$^\circ$ were observed for different foil thickness and irradiation conditions. The proton beams show a pronounced collimation over the whole energy range. We attribute the small divergence to a steep longitudinal electron density gradient at the target rear in comparison with $\mathrm{\mu m}$ thick target, and the collimation feature is related to an exponentially decaying electron density in transverse dimension. Our interpretation is supported by both an analytical model  and the two-dimensional particle-in-cell (PIC) simulations, revealing new physics that occur in the interaction of high-intensity laser with nanometer thin targets. 

\section{Experiment}

The experiments were performed with the ATLAS Ti:sapphire laser system at Max-Planck-Institute for Quantum Optics. This system delivers pulses with a duration of 30 $fs$ FWHM centered at 795 $\mathrm{nm}$ wavelength. The initial laser contrast is $3\times10^{-6}$ at 2 $ps$ before the peak of main pulse. A re-collimating double plasma mirror system was introduced to further enhance the value to $10^{-9}$. 400 $mJ$ laser energy was delivered on target. A $90^\circ$ $f/2$ off-axis parabolic mirror focuses the pulses to a measured FWHM diameter spot size of 3 $\mathrm{\mu m}$, yielding peak intensity of $8\times10^{19}W/cm^{2}$. DLC foils of thickness 5, 10 and 20 $\mathrm{nm}$ have been irradiated under normal incidence for varying spot size, the actual spot size on target in the range of 3-19.2 $\mathrm{\mu m}$ has been adjusted by moving the target along the laser axis. 
 
\begin{figure}[!ht]
  \includegraphics[width=1.0\columnwidth]{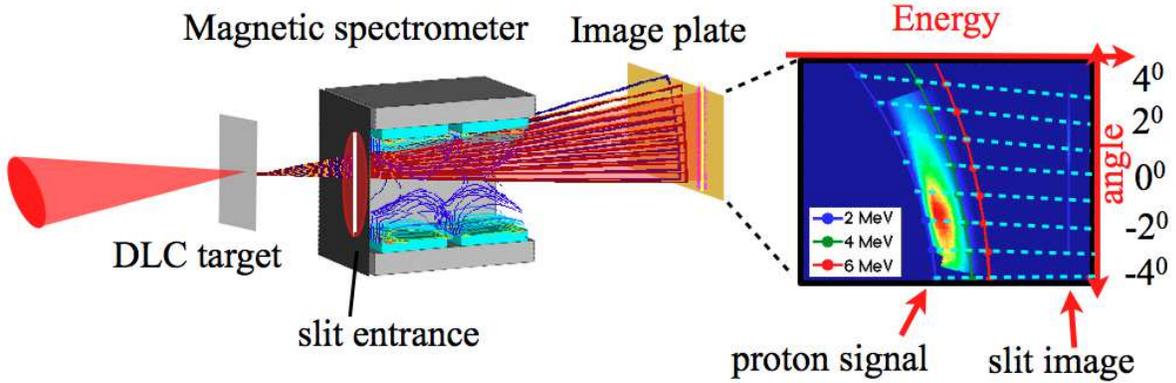}
  \caption{(color online). Experimental setup. The divergence of protons characterized with the magnetic spectrometer and Image plate (IP). The trajectory of protons through the magnetic field as well as the magnetic field structure is shown in the setup picture. The resulting isoenergy contours of the magnetic spectrometer are superimposed with a raw image of proton energy-angular distribution as it appears at the IP. The example is obtained with a 10 $\mathrm{nm}$ target displaced by 100 $\mathrm{\mu m}$ from the laser focal plane. } \label{fig1}
\end{figure}

Fig. \ref{fig1} shows the magnetic spectrometer with a gap of 14 $\mathrm{cm}$ employed for the proton beam measurements. A long vertical entrance slit of 300 $\mathrm{\mu m}$ width is placed in front of the magnetic field. This configuration enables angularly-resolved high-accuracy energy distribution measurement. Fujifilm BAS-TR image plates (IPs) were positioned at a distance of 30 $\mathrm{cm}$ behind the magnets to capture ion phase space over an angular range of 8$^\circ$. The IPs have been absolutely calibrated at MLL Tandem accelerator \cite{IP}. A Layer of 45 $\mathrm{\mu m}$ Al foil was added in front of the IPs to block heavy ions and to protect IPs from direct and scattered laser light. Protons with energies beyond 2 MeV are recorded. A typical raw image from a 10 $\mathrm{nm}$ DLC foil is shown in Fig. \ref{fig1}. The image of the entrance slit, $i.e.,$ the zero line, and the low energy cutoff line from proton signal allows to extrapolate the average magnetic field for different angles. These are used to transfer the spatial information from raw image to the energy angular distribution of protons.The trajectory of proton beams through the dipole magnets as well as the magnetic field was carefully modeled and calibrated, as shown in Fig. \ref{fig1}. The resulted isoenergy contours of the spectrometer shows how the proton propagate through the spectrometer and form a angular distribution on the detector, the isoenergy curve shows fair agreement of the experimental raw image for 2 MeV low energy cutoff by the Al foil.

\begin{figure}[!ht]
  \includegraphics[width=1.0\columnwidth]{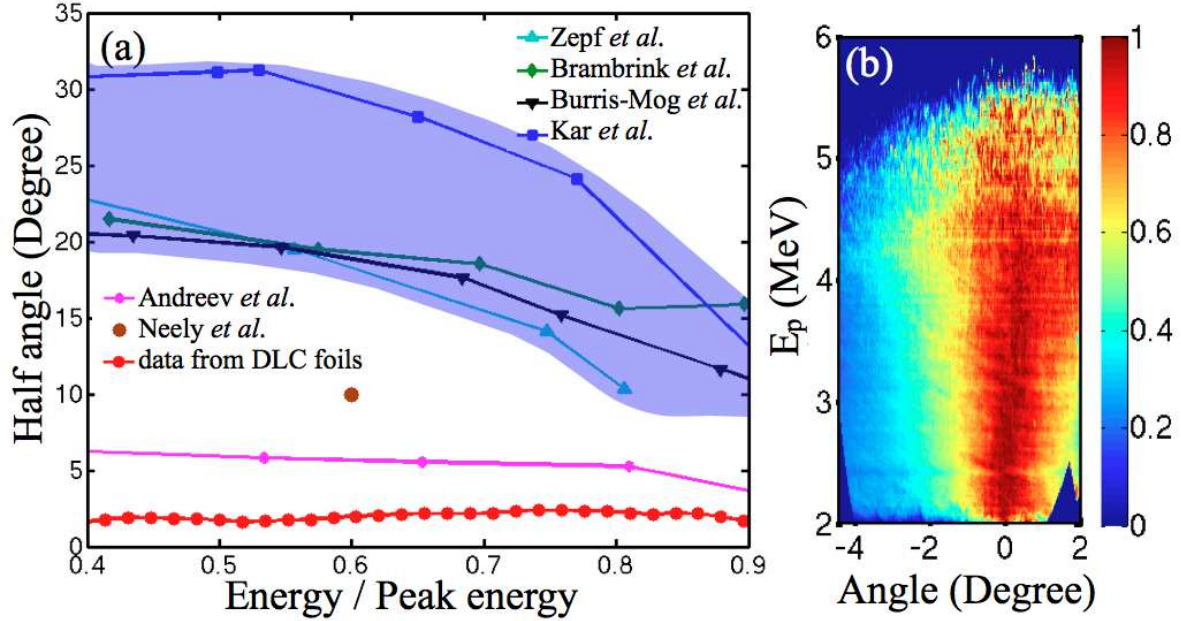}
  \caption{(color online). (a) Beam divergence half angle as a function of proton energy for the most collimated beam from DLC foils, along with other data published in the literature from $\mathrm{\mu m}$ thick targets (the light blue area, presented by blue, green, black and cyan curves) \cite{divergence,solenoid,divcontrol} and from nm scale thin targets (brown dot and magenta curve) \cite{thinfoil2,thinfoil3}. (b) Experimentally processed result of the data presented in Fig. \ref{fig2}(a) after normalization, where the color scale denotes the normalized fluence for given energy.} \label{fig2}
\end{figure}

The smallest divergence was observed with a 10 $\mathrm{nm}$ target displaced by 100 $\mathrm{\mu m}$ from the laser focal plane. Fig. \ref{fig2}(b) shows the energy-angular distribution of the example from Fig. \ref{fig1} after normalization in order to highlight the collimation over the complete energy range. The divergence is almost constant over the detected energy range, resembling the afore mentioned collimation. The angular distribution is fitted by a Gaussian function for each energy value. We define our divergence by the half value of Full-width half-maximum (FWHM) of the fitting profile and plot these values as a function of proton energy normalized to the peak energy, as shown in Fig. \ref{fig2}(a) by the red curve. This enables a comparison to established results \cite{divergence,solenoid,divcontrol} that also presented in Fig. \ref{fig2}(a). Compared to $\mathrm{\mu m}$ thick targets, the half angle from 10 nm foil is reduced by a factor of 10. Moreover, the typical increasing of divergence with decreasing energy is not observed. The results from experiment with 50 nm \cite{thinfoil2} and 800 nm foils \cite{thinfoil3} are included as well, indicating an overall reduction of divergence with decreasing target thickness. 

\begin{figure}[!ht]
  \includegraphics[width=1.0\columnwidth]{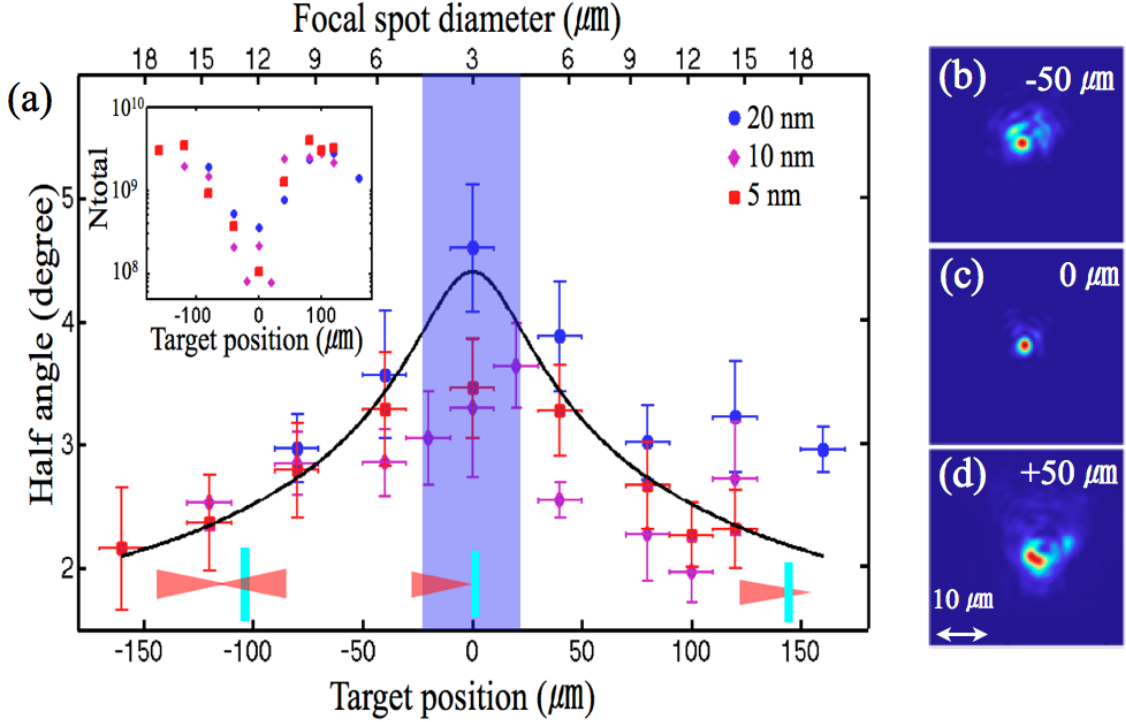}
  \caption{(color online). (a)Beam divergence (half angle) for varying thickness of DLC foils and target positions. The lower axis shows the target position, where +/- means the foils were placed before/after laser focal plane, as indicated by the small pictures. While the upper axis denotes the focal spot FWHM diameter $D_{L}$ given by the spot size of a perfect Gaussian beam in the propagation axis and quantified by the measurements from real laser intensity distribution. Here the vertical error bar indicates the standard deviation for each shot while the horizontal error bar shows the positioning accuracy of about 10 $\mathrm{\mu m}$, smaller than the Rayleigh length of 25 $\mathrm{\mu m}$ marked by the light blue area. The black curve is an empirical fitting curve showing the divergence scale with laser FWHM diameter $(D_{L})^{-1/2}$. The inset presents the corresponding total number of protons for each shot, extracted from the measured proton spectrum above 2 MeV and corresponding divergence from the parameter scan. The Monitored laser intensity distribution is shown at three target positions -50 $\mathrm{\mu m}$ (b), 0 $\mathrm{\mu m}$ (c) and 50 $\mathrm{\mu m}$ (d). } \label{fig3}
\end{figure}

Further on, the thickness of the targets and their positions with respect to the focal plane of the laser were varied. As shown above, the divergence showed no noticeable dependence on energy. Therefore, we plotted the average value of the half angle as a function of the target position in Fig. \ref{fig3}(a). The divergence is maximized with a value of 4.6$^\circ$ in the focus plane for the thickest foil of 20 nm. For thinner foils (5 and 10 nm), these values are reduced to 3.3$^\circ$. The divergence of the protons decreases with increasing focal spot size on the target when moving the target out of focal plane to both sides beyond the Rayleigh length. The smallest divergence was obtained in the target position of +100 $\mathrm{\mu m}$, which is our example of Fig. \ref{fig2}(a),(b). Additionally, the total number of protons above 2 MeV even increases when enlarging the target positions, i.e. when the divergence reduces, as shown in the inset of Fig.  \ref{fig3}(a). Note that the spectral distributions remained exponential throughout the whole parameter scan, in contrast to the case of the previous work with circular polarization\cite{thinfoil, peakedenergy1} or under much higher laser intensity with linear polarization from experiments\cite{peakedenergy2}.

\section{Discussion and outlook}

To explain our results, we derive the emission angle of ions $\theta=\arctan(\int^{\infty}_{0} E_{y} dt/\int^{\infty}_{0} E_{z} dt)$  based on a quasi-stationary electrostatic model, where $y$ and $z$ are the transversal and the longitudinal dimension, respectively. The electric field strengths are determined by $E=-\nabla\Phi$. With a Boltzmann distribution of the hot electrons, the electric field is then given by $E_{y}\propto-\case{k_{B}T_{e}}{e}\case{\partial n_{e}}{\partial y}$, $E_{z}\propto-\case{k_{B}T_{e}}{e}\case{\partial n_{e}}{\partial z}$, where $n_{e}$ is the electron density. Defining the local field direction as $\alpha=E_{y}/E_{z}$, the emission angle $\theta=\arctan({\int^{\infty}_{0} \alpha E_{z} dt}/{\int^{\infty}_{0} E_{z} dt})\approx\arctan$$<$$\alpha$$>$$=\arctan$$<$${\case{\partial n_{e}}{\partial y}}/{\case{\partial n_{e}}{\partial z}}$$>$, where the angle bracket denotes the average along the ion trajectory. We assume the electron density $n_{e}=n_{0}\cdot \xi(y)\cdot exp[-\case{z}{l_{0}}]$ with a transverse profile $\xi(y)$ and a longitudinal exponential distribution, as widely accepted \cite{scalelength}. Here $n_{0}$ is a constant denoting the electron density, and $l_{0}$ is the longitudinal density scale length. One derives
\begin{equation}
\theta= \arctan<-\case{\case{\partial \xi(y)}{\partial y}}{\xi(y)}\cdot l_{0}>, \label{eq1}
\end{equation}
The emission angle $\theta$ represents the characteristic value of the measured divergence.

Eq. (\ref{eq1}) shows how the divergence relies on the electron density distribution. It not only predicts the influence of transverse electron density on the divergence of ions, consistent with \cite{elespatial}, but highlights the importance of longitudinal electron density distribution as well. This fact has not been well clarified so far. A steep longitudinal electron density gradient, i.e., small $l_{0}$, will reduce the divergence of ions.

\begin{figure}[!ht]
  \includegraphics[width=1.0\columnwidth]{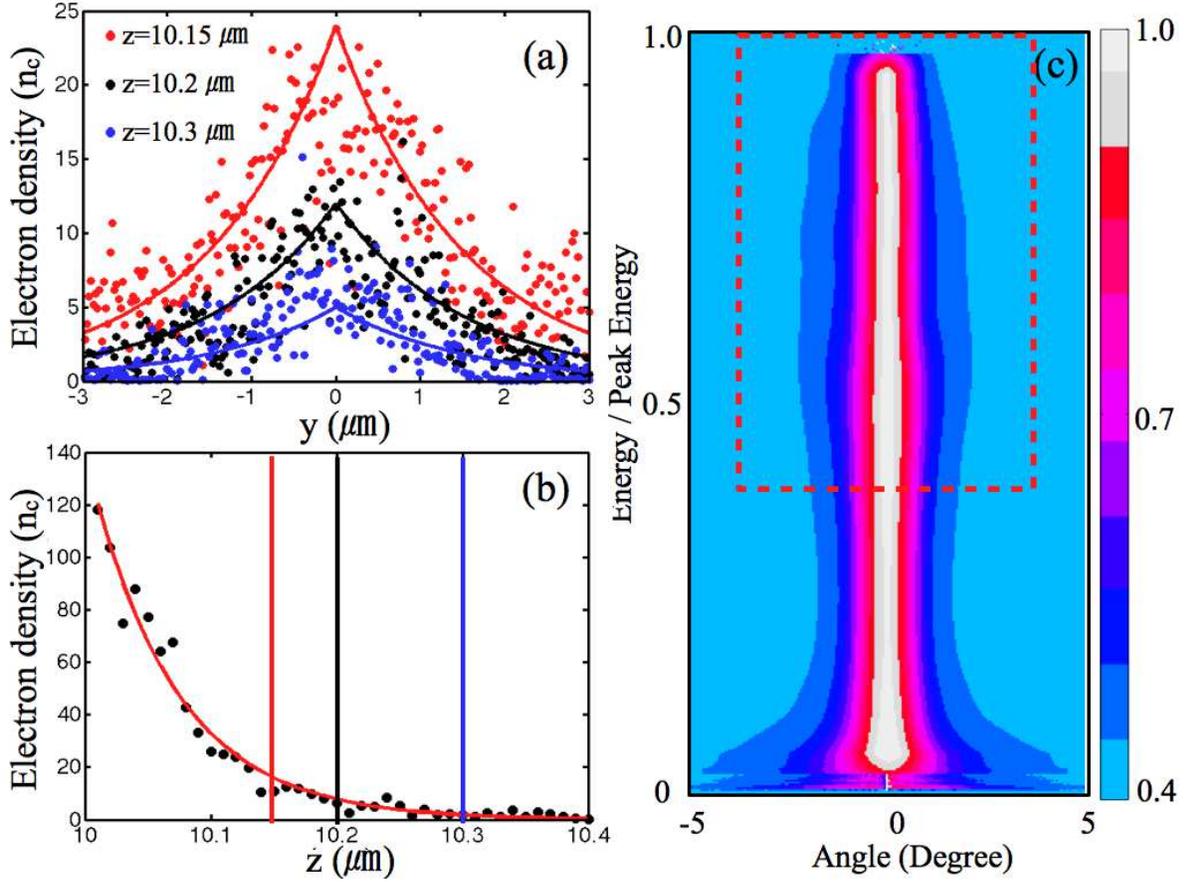}
  \caption{(color online). (a) Transverse electron density distribution ($y$ axis) at three different $z$ positions (as denoted by different color lines in Fig. 4(b)) and (b) Longitudinal electron density distribution ($z$ axis) at $y=0$ at $t=30T$ (when the laser pulse has just left the target),  where $T$ corresponds to one laser cycle. Those distributions are obtained with 40 nm target irradiated at the best focus position. The front side of the target was initially located at $z=10$ $\mathrm{\mu m}$. The dots denote the simulated results while the curves are the exponential functions. (c)Simulation result for proton angular distribution at $t=100T$ after normalization as a comparison to the experiment result in Fig. \ref{fig2}(b). Here the red rectangular marks the experimental observation window. } \label{fig4}
\end{figure}

The scale length $l_{0}$ can be estimated as the local Debye length $\lambda_{D}=(\epsilon_{0}k_{B}T_{e}/n_{0}e^{2})^{1/2}$\cite{plasmaexp}, where $k_{B}T_{e}$ is the electron mean energy. For a given $k_{B}T_{e}$, $l_{0}$ is inversely proportional to the square root of the electron density $n_{0}$. In the case of $\mathrm{\mu m}$ thick targets, due to the large angular spread of few 10s of degrees for hot electrons inside the target \cite{eleangle}, $n_{0}$ drops significantly at the target rear, resulting in a typical $l_{0}$ of few $\mathrm{\mu m}$ \cite{scalelength}. When the target thickness is reduced from tens of $\mathrm{\mu m}$ to submicrometer, the influence from electron propagation through the target is suppressed and the angular spread of hot electrons is substantially reduced, altogether resulting in a higher $n_{0}$. Moreover, with the reduction of thickness, the recirculation of the hot electrons is enhanced \cite{highden}, which can further increase $n_{0}$. Those arguments imply that a small $l_{0}$ exists at the rear side of thinner targets, which in turn resulting in a small divergence of ions. When the thickness of the target is down to $\mathrm{nm}$ scale, an additional factor arises; The pondermotive force pushes a large portion of electrons away from the ions, which further increase the electron density as well as the density gradient, in turn reducing $l_{0}$. Ultimately in light sail regime, a balance between ponderomotive force and charge separation field is formed, and a dense electron layer pushed by the laser pulse drives the ion acceleration, the acceleration field is confined between two oppositely charged plates within 10s of nm \cite{denelec,klap}.

In previous experiments, the transverse electron density was found to be well approximated by a bell shape profile $\xi(y)\propto exp(-4\ln2\case{y^{2}}{D^{2}})$ for $\mathrm{\mu m}$ thick target \cite{elespatial,bellshape}, where $D$ corresponds to the FWHM diameter of the transverse Gaussian profile. Note that typically $D=D_{L}+2d\tan\beta$ is assumed \cite{diameter}, which is much bigger than the FWHM diameter of laser spot $D_{L}$ due to the $\mathrm{\mu m}$ target thickness $d$  and large half angle of the electron angular spread $\beta$. In this case Eq. (\ref{eq1}) reads $\theta\propto\arctan$$<$$8\ln2\case{y}{D^{2}}\cdot l_{0}$$>$.  It is monotonically increasing with local transverse position $y$. Since the high acceleration field appears in the centre, the high-energy ions originate from small $y$. This implies a reduction of divergence with increasing energy, consistent with \cite{divergence,solenoid,divcontrol,divsum} but contradicts our results. Regarding Eq. (\ref{eq1}), the collimation manner observed in our experiments can be explained by an exponentially decaying transverse density distribution $\xi(y)\propto exp[-\case{|y|}{l_{y}}]$, where $l_{y}$ is the transverse density scale length. With such a given transverse profile, a constant value of $\theta\propto\arctan(l_{0}/l_{y})$,independent on the energy and the trajectory of ions, is obtained, which is consistent with our observations. For ultrathin targets, much more low-energetic electrons can penetrate through the nm foil and contribute to the acceleration field. Such low-energy electrons are trapped close to target and reach equilibrium in a short time and form a Boltzmann distribution along the transverse dimension as well, leading to a constant divergence as observed in the experiment. 

To verify our arguments and analytical model, 2D PIC simulation were performed with the KLAP code \cite{klap}. Solid density ($n_0=350n_c$, where $n_c$ is the critical density) plasma slab of 40 nm thickness was considered. The initial temperature of electrons is 1 keV. The simulation box is 60 $\lambda$ in laser direction ($x$) and 20 $\lambda$ in transverse direction ($y$) in 2D with a resolution of 200 cells/$\lambda$ and 40 cells/$\lambda$, respectively. Each cell is filled with 400 quasiparticles.  A linearly polarized laser pulse with a Gaussian envelope in both the spatial  and temporal distribution with a FWHM diameter  $D_{L}$ of 3 $\mathrm{\mu m}$ and a FWHM duration of 33 fs, is used to replicate the experiment conditions.

Fig. \ref{fig4}(a) shows the transverse electron density profile at three different $z$ positions behind a 20 nm foil. Indeed, the profiles are found to be well represented by an exponentially decaying profile with a scale length $l_{y}$, similar to the radius of laser spot $D_{L}/2$ of approximately 1.5 $\mathrm{\mu m}$. [see Fig. \ref{fig4}(a)]. The longitudinal density distribution from the simulation in Fig. \ref{fig4}(b) is represented by an exponential function as well. The density scale length $l_{0}=$1/12 $\mathrm{\mu m}$, is much smaller than the typical scale length $l_{0}$ for $\mathrm{\mu m}$ targets. Such an electron density distribution finally leads to a well collimated proton beam[see Fig. \ref{fig4}(c)], in a good agreement with the experimental observation in Fig. \ref{fig2}(b). The divergence does not depend on energy, with a constant value of about 2$^\circ$ over the whole energy range. By using $l_{y}$ and $l_{0}$ from simulation, Eq. (\ref{eq1}) gives a half angle of 3.2$^\circ$.  

Furthermore, Eq. (\ref{eq1}) predicts the reduction of $\theta$ with increasing the transverse scale length of electron distribution $l_{y}$, which in the case of ultrathin foils, is related to the laser FWHM diameter $D_{L}$, as confirmed by our target position scan[cf. Fig. \ref{fig3}(a)]. Although, a precise determination of the values of $l_{y}$ and $l_{0}$ accordingly to different $D_{L}$ is beyond the capacity of our simple model and our laser intensity distribution outside the focal plane is not perfect Gaussian, we found that the divergence half angle roughly scales with $(D_{L})^{-1/2}$ experimentally, i.e., $l_{0}/l_{y}\propto (D_{L})^{-1/2}$. Note that the laser intensity distribution was monitored for different target positions during the experiment campaign. The laser intensity distribution is close to an uniform Gaussian profile for most of the target positions, as shown in Fig. \ref{fig3}(a), (b) and (d). We therefore ignore the influence of possible nonuniformity of the laser intensity distribution on the divergence. 

\begin{figure}[!ht]
  \includegraphics[width=1.0\columnwidth]{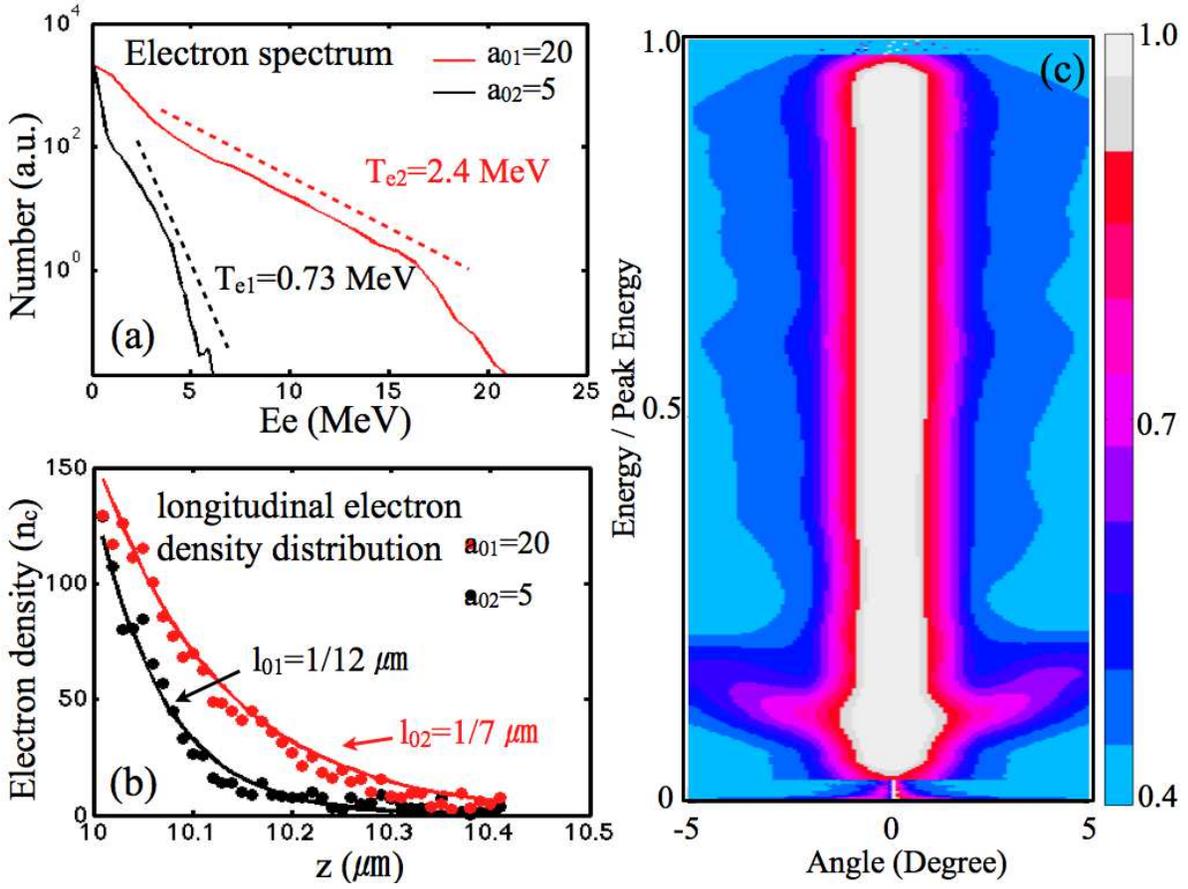}
  \caption{(color online). (a) Electron spectrum at $t=30T$ and (b) Corresponding longitudinal electron density distribution ($z$ axis) at $y=0$ at $t=30T$ (when the laser pulse has just left the target) with two different laser intensities $a_{0}=20$ (red curve) and  $a_{0}=5$ (black curve), where $T$ corresponds to one laser cycle. The front side of the target was initially located at $z=10$ $\mathrm{\mu m}$. (c)Simulation result for proton angular distribution at $t=100T$ with $a_{0}=20$.} \label{fig5}
\end{figure}

Our experiment were carried out with a small laser system. To investigate the effect of larger laser intensity, further simulations were carried out with identical parameters except for a large $a_{0}=20$ as compared to our experimental conditions with $a_{0}=5$, where the laser intensity $I_{0}$ is 16 times bigger as $a_{0}$ is 4 times bigger.
The simulation results are shown in Fig. \ref{fig5}, the electron temperature is increased by more than 3 (from 0.73 MeV to 2.4 MeV) with higher laser intensity ($I_{0}$ is 16 times bigger). Our model predicts that the divergence depends on $l_{0}$, while $l_{0}$ increases roughly with the square root of electron temperature, i.e., from 1/12 $\mathrm{\mu m}$ to 1/7 $\mathrm{\mu m}$ in our example. Therefore we expect a two times larger divergence angle, consistent with our simple model. Still, the ion beams exhibit an almost constant divergence angle under higher laser intensity. Note that the result of proton angular distribution from simulation in Fig. \ref{fig5} was normalized to the maximum proton energy, where the maximum proton energy is increased from 6 MeV to roughly 30 MeV by a factor of 5 with higher laser intensity. Our simulation indicates that the divergence depends only weakly on the laser intensity with other parameters unchanged, i.e., $\theta\propto (I_{0})^{1/4}$.
 
\section{Conclusion}

In conclusion, we investigated the divergence of proton beams generated from ultrathin DLC foils in detail. We demonstrated well collimated proton beams with divergence half angle as low as 2$^\circ$. This constitutes the smallest value reported so far and one order of magnitude lower than achieved with $\mathrm{\mu m}$ targets. As a consequence, 100 times increase in proton fluence is observed \cite{cell}. Moreover, the proton beams are well-collimated over the complete energy range and can be further optimized by adjusting the focal spot size. Those ultrasmall divergence and well collimation is considered as an intrinsic characteristic of $\mathrm{nm}$ ultrathin foils. We expect these observations to be of particularly interest for applications. For example, the divergence angle is sufficiently small to enable lossless beam transport utilizing conventional ion optics such as quadrupoles with typically a few $msr$ acceptance angle \cite{qp}. More directly, investigations that rely on high proton flux, can benefit from the ultrasmall divergence, possibly enhanced by focusing laser accelerated ions with shaped $\mathrm{nm}$ targets \cite{shapenp}, a development that is pursued in our laboratory.

\begin{ack}
This work was supported by DFG through Transregio SFB TR18 and the DFG Cluster of Excellence Munich Centre for Advanced Photonics (MAP) and the Association EURATOM -Max-Planck-Institut f$\ddot{u}$r Plasmaphysik. J.H. Bin, D. Kiefer acknowledge financial support from IMPRS-APS.
\end{ack}


\section*{References}

\begin{thebibliography}{99}

\bibitem{experiments} Maksimchuk A {\it et al}., 2000 {\it Phys. Rev. Lett. } {\bf 84} 4108;
                      Clark E L {\it et al}., 2000 {\it Phys. Rev. Lett.} {\bf 85} 1654;
                      Snavely R A {\it et al}., 2000 {\it Phys. Rev. Lett.} {\bf 85} 2945;
                      Hatchett S P {\it et al}., 2000 {\it Phys. Plasmas} {\bf 7} 2076;
                      Daido H {\it et al}., 2012 {\it Rep. Prog. Phys.} {\bf 75} 056401.                   
                                                
 \bibitem{application}Roth M {\it et al}., 2001 {\it Phys. Rev. Lett.} {\bf 86} 436;
                      Malka V {\it et al}., 2004 {\it Med. Phys.} {\bf 31} 1587;
                      Krushelnik K {\it et al}., 2000 {\it IEEE Trans. Plasma Sci.} {\bf 28} 1184;
                      Cobble J A {\it et al}., 2002 {\it J. Appl. Phys.} {\bf 92} 1775;
                      Habs D {\it et al}., 2009 {\it Eur. Phys. J.} D {\bf 55} 279.
                                          
\bibitem{divergence}Zepf M {\it et al}., 2003 {\it Phys. Rev. Lett.} {\bf 90} 064801;
                                          Brambrink E {\it et al}., 2006 {\it Phys. Rev. Lett.} {\bf 96} 154801.
                                                                                    
\bibitem{divcontrol}Kar S {\it et al}., 2011 {\it Phys. Rev. Lett.} {\bf 106} 225003.
 
\bibitem{qp}Schollmeier M {\it et al}., 2008 {\it Phys. Rev. Lett.} {\bf 101} 055004.
                                    
\bibitem{solenoid}Burris-Mog T {\it et al}., 2011 {\it Phys. Rev. ST Accel. Beams} {\bf 14} 121301.

 \bibitem{secondp}Toncian T {\it et al}., 2006 {\it Science} {\bf 312} 410.

\bibitem{lenstarget}Kar S {\it et al}., 2008 {\it Phys. Rev. Lett.} {\bf 100} 105004.

\bibitem{droplets}Sokollik T {\it et al}., 2009 {\it Phys. Rev. Lett.} {\bf 103} 135003.

\bibitem{curved}Patel P K {\it et al}., 2003 {\it Phys. Rev. Lett.} {\bf 91} 125004.

\bibitem{tnsa}Wilks S C {\it et al}., 2001 {\it Phys. Plasmas} {\bf 8} 542.

\bibitem{elespatial}Fuchs J {\it et al}., 2003 {\it Phys. Rev. Lett.} {\bf 91} 255002;
                                      Schollmeier M {\it et al}., 2008 {\it Phys. Plasmas} {\bf 15} 053101.
                                   
\bibitem{thinfoil}Henig A {\it et al}., 2009 {\it Phys. Rev. Lett.} {\bf 103} 045002;
                               Henig A {\it et al}., 2009 {\it Phys. Rev. Lett.} {\bf 103} 245003; 
                              Steinke S {\it et al}., 2010 {\it Laser Part. Beams} {\bf 28} 215.
                               
\bibitem{thinfoil2}Neely D {\it et al}., 2006 {\it Appl. Rev. Lett.} {\bf 89} 021502.

\bibitem{thinfoil3}Andreev A {\it et al}., 2010 {\it New J. Phys.} {\bf 12} 045007.
                                                                 
\bibitem{dlc}Ma W {\it et al}., 2011 {\it Methods Phys. Res.} A {\bf 655} 53.

\bibitem{IP}Reinhardt S, 2012  {\it Ph. D. thesis} (Ludwig-Maximilians-Universit$\ddot{a}$t M$\ddot{u}$nchen, Munich).

\bibitem{peakedenergy1}Robinson A P L {\it et al}., 2008 {\it New J. Phys.} {\bf 10} 013021.

\bibitem{peakedenergy2}Kar S {\it et al}., 2012 {\it Phys. Rev. Lett.} {\bf 109} 185006.

\bibitem{scalelength}Mackinnon A J {\it et al}., 2001 {\it Phys. Rev. Lett.} {\bf 86} 1769;
                                         J$\ddot{a}$ckel  O J {\it et al}., 2010 {\it New J. Phys.} {\bf 12} 103027;
                                         Fuchs J {\it et al}., 2007 {\it Phys. Rev. Lett.} {\bf 99} 015002.

\bibitem{plasmaexp}Mora P, 2003 {\it Phys. Rev. Lett.} {\bf 90} 185002.

\bibitem{eleangle}Santos J J {\it et al}., 2002 {\it Phys. Rev. Lett.} {\bf 89} 025001.

\bibitem{highden}Mackinnon A J {\it et al}., 2002 {\it Phys. Rev. Lett.} {\bf 88} 215006.

\bibitem{denelec}Macchi A {\it et al}., 2005 {\it Phys. Rev. Lett.} {\bf 94} 165003;
                                   Robinson A P L {\it et al}., 2009 {\it Plasma Phys. Control. Fusion} {\bf 51} 024004.

\bibitem{klap}Yan X Q {\it et al}., 2008 {\it Phys. Rev. Lett.} {\bf 100} 135003.

\bibitem{diameter}Schreiber J {\it et al}., 2006 {\it Phys. Rev. Lett.} {\bf 97} 045005.

\bibitem{bellshape}Romagnani L {\it et al}., 2005 {\it Phys. Rev. Lett.} {\bf 95} 195001.

\bibitem{divsum}Roth M {\it et al}., 2005 {\it Plasma Phys. Control. Fusion} {\bf 47} B841.

\bibitem{cell}Bin J {\it et al}., 2012 {\it Appl. Phys. Lett.} {\bf 101} 243701.

\bibitem{shapenp}Bartal T {\it et al}., 2011 {\it Nature Phys.} {\bf 8} 139.

\end{thebibliography}
\end{document}